\PassOptionsToPackage{dvipsnames}{xcolor}
\documentclass[10pt,conference]{IEEEtran}
\IEEEoverridecommandlockouts
\usepackage{cite}
\usepackage{amsmath,amssymb,amsfonts}
\usepackage{algorithmic}
\usepackage{graphicx}
\usepackage{textcomp}
\usepackage{xcolor}
\usepackage{balance}
\usepackage[all]{nowidow}

\usepackage[hidelinks]{hyperref}
\usepackage[capitalise]{cleveref}

\usepackage{fancyhdr}
\begin{document}

\title{Towards an Optimized Benchmarking Platform for CI/CD Pipelines %
    \thanks{Partially funded by the Bundesministerium f{\"u}r Forschung, Technologie und Raumfahrt (BMFTR, German Federal Ministry of Research, Technology and Space) in the scope of the Software Campus 3.0 (Technische Universit\"at Berlin) program -- 01IS23068.\\\\
    © 2025 IEEE.
    Personal use of this material is permitted.
    Permission from IEEE must be obtained for all other uses, in any current or future media, including reprinting/republishing this material for advertising or promotional purposes, creating new collective works, for resale or redistribution to servers or lists, or reuse of any copyrighted component of this work in other works.\\
    DOI: \href{https://doi.org/10.1109/IC2E65552.2025.00010}{10.1109/IC2E65552.2025.00010}
    }
}

\author{\IEEEauthorblockN{Nils Japke\IEEEauthorrefmark{1}, Sebastian Koch\IEEEauthorrefmark{1}, Helmut Lukasczyk\IEEEauthorrefmark{2}, David Bermbach\IEEEauthorrefmark{1}}
    \IEEEauthorblockA{\IEEEauthorrefmark{1}\textit{Technische Universit\"at Berlin}\\
        \textit{Scalable Software Systems Research Group}\\
        \{nj,seko,db\}@3s.tu-berlin.de}
    \IEEEauthorblockA{\IEEEauthorrefmark{2}\textit{DATEV eG}\\
        Helmut.Lukasczyk@DATEV.DE}
}

\maketitle
\thispagestyle{plain} %

\begin{abstract}
    Performance regressions in large-scale software systems can lead to substantial resource inefficiencies, making their early detection critical.
    Frequent benchmarking is essential for identifying these regressions and maintaining service-level agreements (SLAs).
    Performance benchmarks, however, are resource-intensive and time-consuming, which is a major challenge for integration into Continuous Integration / Continuous Deployment (CI/CD) pipelines.
    Although numerous benchmark optimization techniques have been proposed to accelerate benchmark execution, there is currently no practical system that integrates these optimizations seamlessly into real-world CI/CD pipelines.
    In this vision paper, we argue that the field of benchmark optimization remains under-explored in key areas that hinder its broader adoption.
    We identify three central challenges to enabling frequent and efficient benchmarking: (a) the composability of benchmark optimization strategies, (b) automated evaluation of benchmarking results, and (c) the usability and complexity of applying these strategies as part of CI/CD systems in practice.
    We also introduce a conceptual cloud-based benchmarking framework handling these challenges transparently.
    By presenting these open problems, we aim to stimulate research toward making performance regression detection in CI/CD systems more practical and effective.
\end{abstract}


\section{Introduction}
\label{sec:introduction}
In modern large-scale, cloud-based software systems, it is important to catch performance regressions early, since even tiny performance regressions can end up wasting a large amount of compute resources.
In order to detect these performance regressions, software developers need to frequently run benchmarks to evaluate how the performance of the software evolves~\cite{book_bermbach2017_cloud_service_benchmarking,GrambowLehmannBermbach2019,waller_including_2015}.
This is often necessary to validate non-functional requirements according to ISO/IEC~25010~\cite{iso25010}, to ensure service level agreements (SLAs) are met, and aid in user conversion and retention.
Software benchmarks are expensive to execute, however, as they need a dedicated testing environment, which needs to mirror the production environment, and often run for several hours to ensure statistically stable and relevant results~\cite{GrambowUMBS}.

Ideally, benchmarks should run frequently to give software developers insight into their software, e.g., running on every new version, or even commit, as part of a Continuous Integration / Continuous Deployment (CI/CD) pipeline, however, running performance tests this frequently is currently prohibitively expensive.
In order to address this problem, several approaches for minimizing the execution time of benchmarks have been proposed~\cite{japke2025muoptime,Grambow2021,GrambowUMBS,laaber_dynamically_2020,laaber_applying_2021,laaber_mobp_2022,traini_ai-driven_2024,bulej2019initial,bulej2020duet,georges_statistically_2007,he_statistics_2019,metior,AlGhamdi2016,AlGhamdi2020,chen_perfjit_2020}.
Currently, no system exists, however, which can run such benchmark optimizations in a real-world CI/CD pipeline and which is usable for practitioners.
This suggests that (a) using benchmark optimizations is difficult in itself, and (b) even with optimizations, benchmarks still take too long to execute inside a frequently running pipeline.
Despite these challenges, there is growing interest in industry and research to enable software benchmarking inside CI/CD pipelines~\cite{GrambowLehmannBermbach2019,ingo_automated_2020,daly_creating_2021,daly_industry_2019,waller_including_2015}.

We argue that the field of benchmarking optimization strategies for reducing execution time is still under-explored and leaves open key challenges, which can make highly frequent software benchmarking inside pipelines feasible.
These key challenges, as we see them, are (a) the composability of benchmark optimization strategies, (b) automated evaluation of benchmarking results, and (c) usability and complexity issues of benchmark optimization strategies.
In this regard, we envision, and here propose, a conceptual integrated cloud-based benchmarking framework, which addresses these challenges.

In \cref{sec:backround_rel_work}, we start by introducing background and related work.
In \cref{sec:vision}, we present the open research challenges in benchmark optimization strategies listed above.
In \cref{sec:system}, we present our vision of an integrated benchmarking framework targeting these challenges.
Finally, in \cref{sec:conclusion}, we conclude the paper.

\section{Background \& Related Work}
\label{sec:backround_rel_work}
In this section, we define and introduce the terminology and background used throughout the paper, and refer to relevant related work.

\begin{figure}[tbp]
    \centering
    \includegraphics[width=0.2\textwidth]{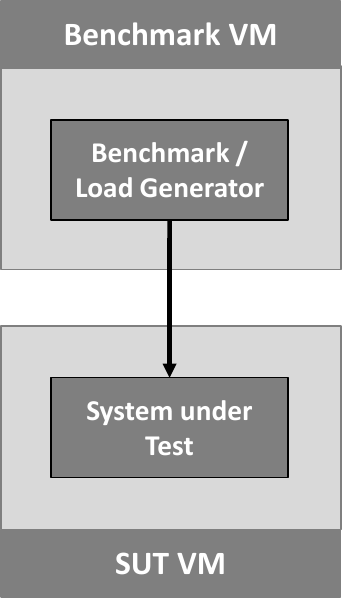}
    \caption{A typical cloud benchmarking scenario: The SUT and the corresponding load generator of the application benchmark are deployed on different VMs in the cloud.
    This ensures that the load generator does not take CPU time away from the SUT by effectively isolating both components.}
    \label{fig:application-benchmark}
\end{figure}

\subsection{Performance Testing in the Cloud}
Nowadays, performance testing often involves cloud infrastructure, because (i) it is simple to set up a short-lived testing environment in the cloud, and (ii) often the software itself is deployed in a cloud environment in production.
Running software benchmarks in the cloud comes with additional challenges, however, since the cloud infrastructure is subject to performance fluctuations, which introduces additional noise to measurements or even measurement bias~\cite{LeitnerCito2016}.
There are techniques to deal with the additional noise, typically requiring replications of benchmarks across different virtual machines (VMs) to produce repeatable results~\cite{book_bermbach2017_cloud_service_benchmarking,LaaberScheunerLeitner2019,AbediOld,AbediCloud}.

\subsection{Types of Benchmarks}
We distinguish two different types of benchmarks.
The first type is called an \textbf{application benchmark}, which is characterized by evaluating the \emph{Sytem under Test} (SUT) from an external perspective.
To achieve this, the SUT is fully set up, nowadays often in a cloud-based staging environment, and stressed using a load generator from the application benchmark, while measuring different performance metrics~\cite{book_bermbach2017_cloud_service_benchmarking}.
In order to guarantee fair measurements, the environment of the SUT should be isolated from external influences as much as possible, which includes the load generator itself, leading to the common practice of deploying the load generator on a different VM in cloud-based deployments~\cite{book_bermbach2017_cloud_service_benchmarking}.
Since an application benchmark evaluates the performance of the SUT from a client perspective, the measurements directly relate to what users of the application experience in production.
\cref{fig:application-benchmark} shows an example deployment of an application benchmark and the corresponding SUT in the cloud.

\begin{figure}[tbp]
    \centering
    \includegraphics[width=0.3\textwidth]{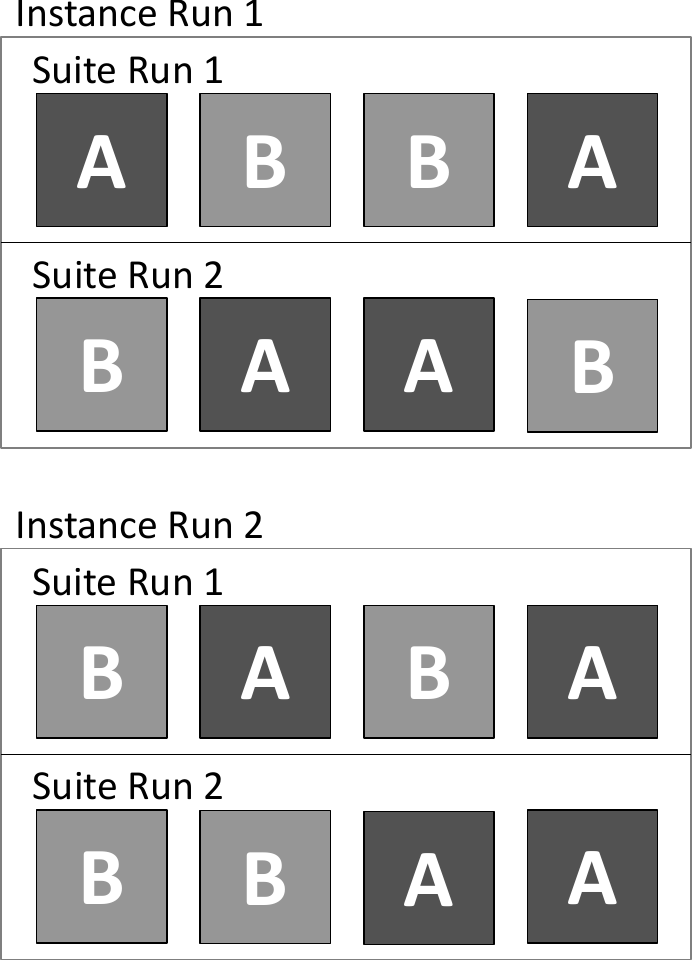}
    \caption{Microbenchmarks with RMIT: Here we see an example of microbenchmarks executed using a simple RMIT schema.
    The microbenchmark \emph{suite} consists of two microbenchmarks, labeled A and B.
    Each microbenchmark is executed for \emph{two iterations}, the \emph{suite} is executed \emph{two times}, and everything is replicated across \emph{two} different cloud \emph{instances} (i.e., VMs).}
    \label{fig:rmit}
\end{figure}

The second type of benchmark we distinguish is called \textbf{microbenchmark}, which measure the performance of a single subroutine (in the following: function) on source code level.
A microbenchmark calls a single function repeatedly with generated input parameters until a predefined time has elapsed, typically one second.
We refer to single function calls as \emph{invocations}, and the entire microbenchmark execution, e.g., repeatedly calling the function for one second, as an \emph{iteration}.
During each invocation, the microbenchmarking framework measures performance characteristics such as execution time and RAM usage of the function and reports an average across all invocations inside an iteration.
Additionally, we refer to the entire collection of microbenchmarks in a software project as the microbenchmark \emph{suite}.
In order to produce repeatable and reproducible results, microbenchmark executions need to be replicated across different components that influence performance measurements, e.g., cloud VMs that might produce different results due to inherent noise~\cite{LeitnerCito2016,paper_bermbach_expect_the_unexpected}.
The current state of the art for running microbenchmarks in cloud environments uses a technique called \emph{Randomized Multiple Interleaved Trials} (RMIT)~\cite{AbediOld,AbediCloud}, which replicates the entire microbenchmark suite across multiple cloud VMs and shuffles repeated iterations of different microbenchmarks.
\cref{fig:rmit} shows an example RMIT schema, where each number can also be configured to a different one.

Microbenchmarks are individually fast to execute, but executing the entire microbenchmark suite of an SUT with high code coverage can become expensive -- in experiments, we have seen microbenchmark suites that take up to 100 hours to run.
Furthermore, microbenchmarks only evaluate small parts of the SUT in isolation, so even an entire suite cannot evaluate the SUT holistically, as complex interactions of functions cannot be represented by microbenchmark results, e.g., one function might become slower, but is called less often as a result.
The main advantage in microbenchmarks is that they can identify even tiny performance regressions, which become masked by generally longer execution times in application benchmarks, and generally find them earlier~\cite{japke2024early}.
They also help to identify root causes due to their narrow scope, as they identify individual functions with performance regressions.

Overall, application benchmarks are complex in setup and can be expensive with typical execution durations below one hour.
In exchange, they can reliably detect performance regressions that matter in practice~\cite{book_bermbach2017_cloud_service_benchmarking,GrambowUMBS}.
Microbenchmarks, in contrast, are simple to setup and run (though the distributed RMIT configuration may be difficult), with execution durations in the tens of hours.
Microbenchmarks tend to detect all performance regressions, not only those that matter in practice, and do so early.
They are a powerful mechanism for pinpointing root causes but tend towards false positive detection of performance regression~\cite{GrambowUMBS}.

\subsection{Benchmark Optimizations}
Many different benchmark optimizations have been proposed in scientific literature.
Some only apply to either application benchmarks or microbenchmarks, but many can be adapted from one type of benchmark to the other.

There are approaches that target a reduction of the number of test cases, in particular selecting a subset of microbenchmarks~\cite{Grambow2021,GrambowUMBS,de_oliveira_perphecy_2017,chen_perfjit_2020}, or reduce the execution time of individual tests~\cite{japke2025muoptime,laaber_dynamically_2020,traini_ai-driven_2024,he_statistics_2019,metior,AlGhamdi2016,AlGhamdi2020}.
Other approaches apply test case prioritization to microbenchmarks~\cite{laaber_applying_2021,laaber_mobp_2022}, which rank microbenchmarks by their likelihood of uncovering performance regressions.
This potentially saves time, when the microbenchmarks fail compared to an acceptable performance threshold in a regression testing scenario.
In such a scenario, new code is rejected, whenever a measured performance regression is higher than the acceptable threshold.
Should a single microbenchmark fail, we do not need to execute any other microbenchmarks.
Therefore, using test case prioritization techniques increases the amount of microbenchmarks that are skipped, whenever a release fails due to performance regressions.
Finally, there are approaches for reducing the need for replicating experiments across different cloud VMs~\cite{bulej2019initial,bulej2020duet}, resulting in infrastructure cost savings, and approaches for exploiting FaaS parallelism~\cite{schirmer2024elastibench}.

\section{Optimizing for Continuous Benchmarking}
\label{sec:vision}
In this Section, we outline our vision for different research directions, in which we see untapped potential.
These all relate to the use of benchmark optimization strategies in CI/CD pipelines.
For each of those research directions, we outline the current state of the art, gaps to our vision, and the research direction that opens up as a result.
We also sketch out additional research directions, supporting the three fields above.

\begin{figure}[tbp]
    \centering
    \includegraphics[width=0.5\textwidth]{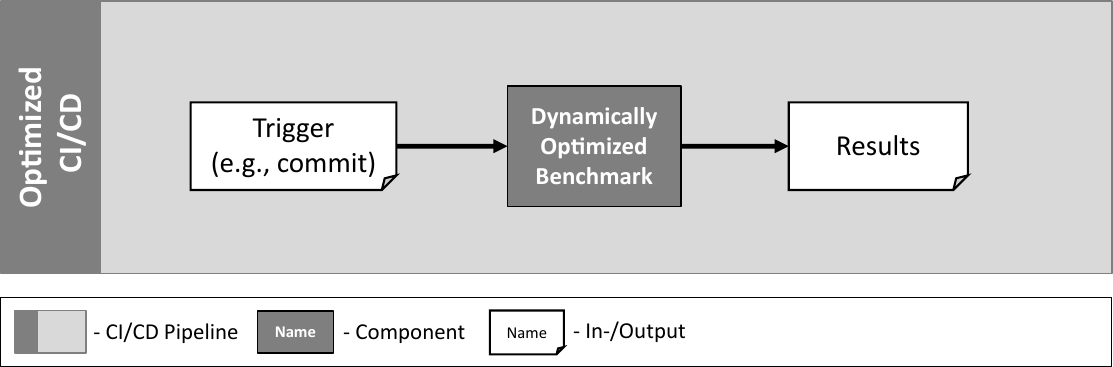}
    \caption{A CI/CD pipeline with a benchmark using dynamic optimization.}
    \label{fig:dynamic-optimization}
\end{figure}

\subsection{Composability of Benchmark Optimizations}
As already discussed in \cref{sec:backround_rel_work}, there is a lot of research on benchmark optimization strategies, which were mainly developed independently of each other.
Many of these optimization strategies can potentially be combined, however, as many of them address different concerns, e.g., first selecting a subset of test cases, then ranking them using test case prioritization, and finally optimizing their runtime.
Yet, it remains unclear what the implications are of using them in combination.
Even within categories, we see potential to combine different approaches, as there are still benchmark optimization strategies that target different parts of benchmarks, e.g., first optimizing warmup time~\cite{laaber_dynamically_2020}, and then optimizing measurement time~\cite{japke2025muoptime}.
Combining many of these techniques has the potential to bring further time savings, but also new challenges, e.g., if both the warmup and measurement phase of a benchmark are shortened using separate approaches, they might affect result quality negatively in combination, but not in isolation.
Still, these compositions of different benchmark optimization techniques present a new research direction with new potential benefits.
We argue that there is much potential for research and, in consequence, for improving CI/CD pipelines.

\subsection{Holistic Evaluation using Real-World CI/CD Systems}
Given that the goal of benchmark optimization strategies is to make benchmarking cheaper, faster, and to enable a higher frequency of benchmarking, the work on composing different benchmark optimization strategies is a natural fit for being integrated into CI/CD pipelines.
While certain benchmark optimizations have been evaluated, whether they can work in CI/CD pipelines~\cite{japke2025muoptime,laaber_applying_2021}, running even the optimized benchmarks still takes too much time to enable frequent use of such pipelines.
This presents a real gap to meet certain needs defined by the industry.

\begin{figure}[tbp]
    \centering
    \includegraphics[width=0.5\textwidth]{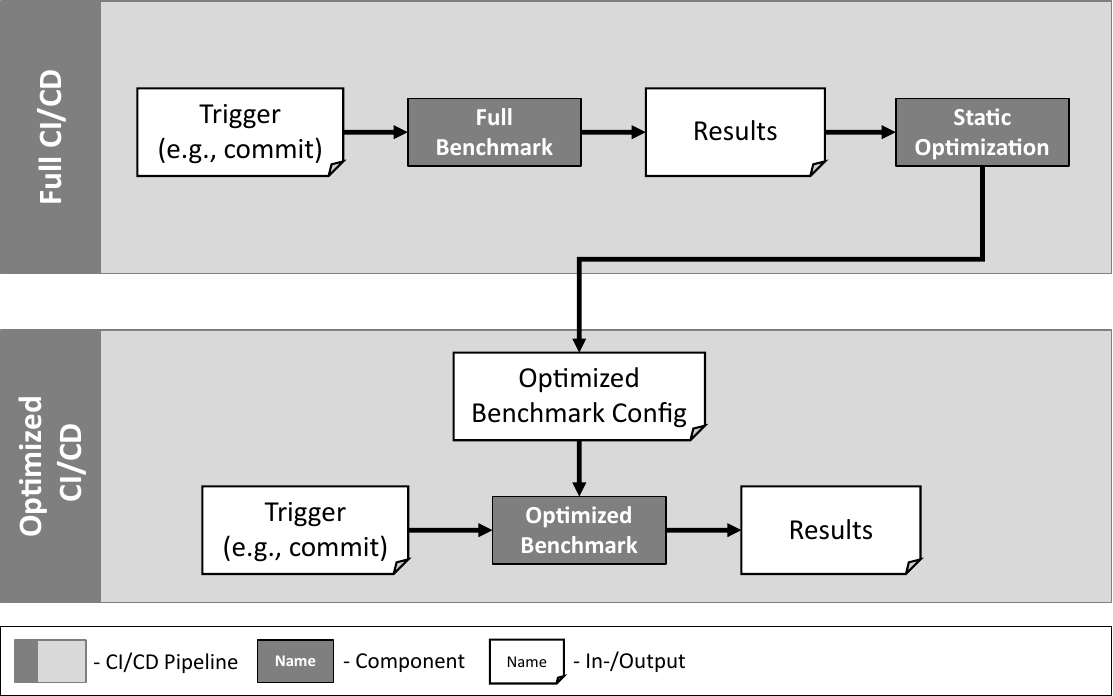}
    \caption{A CI/CD pipeline with a benchmark using static optimization.}
    \label{fig:static-optimization}
\end{figure}

In order to close this gap, different benchmark optimizations need to be evaluated within real CI/CD pipelines over a long period of time.
This provides additional engineering challenges as well.
Some benchmark optimization strategies work dynamically~\cite{laaber_dynamically_2020,he_statistics_2019,metior,AlGhamdi2016,AlGhamdi2020,traini_ai-driven_2024} with additional calculations during a benchmarking run, while others are static~\cite{japke2025muoptime,GrambowUMBS} and need to pre-calculate optimizations using benchmarking results from previous runs.
The static optimization strategies, in particular, only provide benefits on subsequent benchmarking runs, and need to be performed periodically, as the underlying code base obviously changes.
In \cref{fig:dynamic-optimization,fig:static-optimization}, we show two example setups for CI/CD pipelines using dynamic and static optimization approaches.

Realizing such integrated CI/CD pipelines is the first step towards providing a holistic evaluation, whether benchmark optimization is a viable approach for practice.

\subsection{Complexity and Usability of Benchmark Optimizations}
Result analysis is another area, where the complexity increases due to benchmark optimization.
Optimizations that reduce the execution time of benchmarks necessarily also reduce the amount of data we can gather, complicating the analysis.
Certain statistical techniques do not perform well on small sample sizes, such as bootstrap~\cite{Hesterberg2015}, which is often employed for benchmark result analysis~\cite{Kalibera2020,LaaberScheunerLeitner2019,GrambowUMBS}.
This complicates which statistical tools practitioners should choose for optimized benchmarks, especially when using compositions of multiple benchmark optimization strategies.
An integrated benchmarking platform can also aid practitioners in this case, handling the complexity behind the scenes and provide an automated analysis of results, simplifying the interpretation of the results as well.
Therefore, further research into automated analysis techniques needs to provide new tools, which can be embedded into such a platform.
A promising direction in this area is also to explore how emerging AI tools can help with automated result interpretation.

\subsection{Further Research Directions}
There are several other challenges that hinder broad adoption of continuous benchmarking practices as part of CI/CD pipelines.
One is the lack of availability of suitable benchmarks.
For SUTs with standard interfaces, e.g., relational database systems or messaging systems, there is a wealth of existing benchmarks.
In the case of custom application systems, there is very little potential for reusing existing application benchmarks as all interfaces and functionality is application-specific.
While there are first approaches for REST-based microservices~\cite{paper_grambow_benchmarking_microservices,grambow_benchmarking_2020}, application benchmarks usually still have to be written manually.
Likewise, microbenchmark suites are not available in all applications or systems~\cite{Grambow2021} and usually have to be written manually~\cite{RodriguezCancio2016mbgeneration,Jangali2023mbgeneration}:
Writing functional test cases is not the most popular activity for developers and the same holds true for microbenchmarks which are similar code-wise to unit tests.
For both areas, there is potential for AI-generated benchmarks or at least AI-assisted development of benchmarks.

Finally, how best to use both benchmark types in combination is still an open problem.
This is true for interpreting and correlating results but also for the general question of when to run which benchmark type in which configuration.
In this context, proper experiment planning, deployment, and benchmark orchestration still requires lots of human specialist expertise.
Here, research on how to automate these tasks would help towards broad adoption.

\section{Integrated Cloud-Based Benchmarking Framework}
\label{sec:system}
In this Section, we describe our vision of an integrated cloud-based benchmarking framework, resulting from the research opportunities in benchmarking optimizations presented in \cref{sec:vision}.

\subsection{Overview}
\begin{figure}[tbp]
    \centering
    \includegraphics[width=0.45\textwidth]{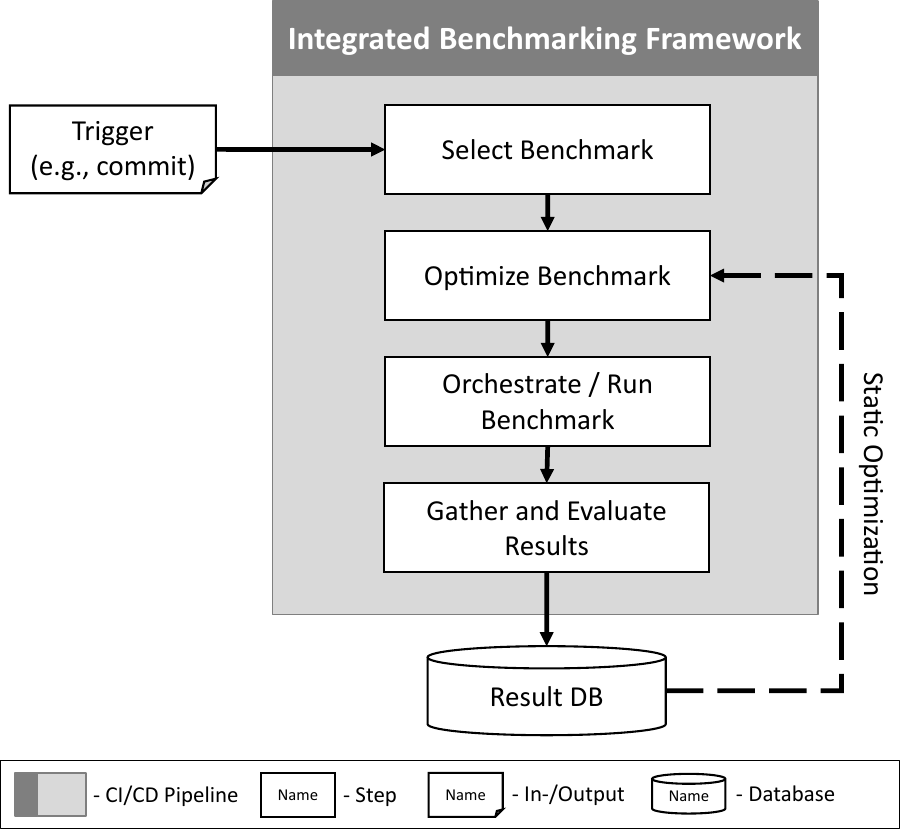}
    \caption{A generic high-level overview of an integrated benchmarking framework for automatically optimizing benchmarks.}
    \label{fig:framework}
\end{figure}

In \cref{fig:framework}, we present a high-level conceptual overview of an integrated framework which runs optimized benchmarks in cloud-based CI/CD pipelines.
Despite the challenges cloud environments present, such a framework running in the cloud profits from easy infrastructure provisioning and deprovisioning for benchmarking purposes.
The framework runs in automated CI/CD pipelines, and integrates (composable) benchmarking optimizations with automated analysis tools to enable rapid benchmarking and result interpretation.

For the purposes of such a framework, we map the benchmarking pipeline to four distinct steps:
\begin{enumerate}
    \item \emph{Select Benchmark}: Select an appropriate benchmark, including different types, such as whether to run microbenchmarks or application benchmarks.
    \item \emph{Optimize Benchmark}: Run benchmark optimizations, dynamic or static, including different types, such as reducing execution time, selecting subsets of test cases, among others.
    \item \emph{Orchestrate / Run Benchmark}: Execute the benchmark using cloud infrastructure with the aforementioned optimizations.
    \item \emph{Gather and Evaluate Results}: Automatically analyze results, potentially aided by new tools and AI for interpretation.
\end{enumerate}
At the end of the pipeline, the framework saves all results to a database, as the data can include important information for static optimizations in subsequent runs.

\subsection{Outlook}
There is potential to also include other research areas than the main direction of benchmark optimizations presented here.
One such area is the automated generation of microbenchmarks~\cite{RodriguezCancio2016mbgeneration,Jangali2023mbgeneration}.
The framework could transparently generate appropriate benchmarks during the first step \emph{Select Benchmark}, which is then further optimized and orchestrated during the rest of the pipeline.

Another research area is relating the results of microbenchmarks to application benchmarks, which has seen some first steps as part of other research in the past~\cite{Grambow2021,GrambowUMBS}.
As the framework encompasses both microbenchmarks and application benchmarks, relating their results to each other can provide further insight into the SUT by showing exactly how the performance of certain functions affects different endpoints.
This could potentially allow for fewer application benchmark runs, as long as microbenchmarks remain in use, as their results provide the necessary information to reason about application-level performance.

\section{Conclusion}
\label{sec:conclusion}
Performance regressions in large-scale, cloud-based software systems can be costly -- they will increase resource consumption and may annoy end users.
Solutions for detecting regressions early usually rely on a benchmarking step as part of CI/CD pipelines~\cite{waller_including_2015,GrambowLehmannBermbach2019}.
In this context, multiple strategies for optimizing benchmark execution have been proposed.

In this vision paper, we argued that more research is needed in this field if we want to make continuous benchmarking broadly available in practice.
Therefore, we described three main research directions in the field of benchmark optimization strategies and gave an overview of auxiliary fields and promising approaches.
We also presented our vision of an integrated cloud-based benchmarking framework, transparently handling the complexity behind using benchmark optimization strategies in practice.

\balance

\bibliographystyle{IEEEtran}
\bibliography{bibliography}

\end{document}